\documentclass[useAMS,usenatbib]{mn2e}
\usepackage{rotating}
\usepackage{journals}

\def\approxgt{\ifmmode \rlap{$>$}{}_{{}_{{}_{\textstyle\sim}}} \else%
$\rlap{$>$}{}_{{}_{{}_{\textstyle\sim}}}$\fi} 
\def\approxlt{\ifmmode \rlap{$<$}{}_{{}_{{}_{\textstyle\sim}}} \else%
$\rlap{$<$}{}_{{}_{{}_{\textstyle\sim}}}$\fi}

\LARGE \normalsize \title[IGR~J00291+5934 in quiescence]{{\it Chandra} observations of the
millisecond X--ray pulsar IGR~J00291+5934 in quiescence}

\author[Jonker et al.]  {P.G.~Jonker$^{1,2}$\thanks{email :
pjonker@cfa.harvard.edu}, S. Campana$^{3}$, D.~Steeghs$^1$, M.A.P.~Torres$^1$,
D.K.~Galloway$^4$, \newauthor C.B.~Markwardt$^{5,6}$, D.~Chakrabarty$^4$, J.~Swank$^5$
\\ $^1$Harvard--Smithsonian 
Center for Astrophysics, 60 Garden Street,
Cambridge, MA~02138, Massachusetts, U.S.A.\\ $^2$SRON, National Institute for
Space Research, Sorbonnelaan 2, 3584~CA, Utrecht, The Netherlands\\
$^3$INAF--Osservatorio Astronomico di Brera, Via Bianchi 46, I-23807 Merate (Lc),
Italy\\ $^4$Center for Space Research, Massachusetts Institute of Technology,
Cambridge, MA 02139, U.S.A. \\ $^{5}$Laboratory for High Energy Astrophysics, NASA, Goddard
Space Flight Center, Greenbelt, MD 20771, U.S.A.\\ $^6$ Department of Astronomy,
University of Maryland, College Park MD 20742, U.S.A. \\}

\begin{document}

\maketitle

\begin{abstract} \noindent In this Paper we report on our analysis of three
{\it Chandra} observations of the accretion--powered millisecond X--ray pulsar
IGR~J00291+5934 obtained during the late stages of the 2004 outburst. We also
report the serendipitous detection of the source in quiescence by ROSAT during
MJD 48830--48839 (July~26--August 4, 1992). The detected 0.3--10 keV source
count rates varied significantly between the {\it Chandra} observations from
$(7.2\pm1.2)\times 10^{-3}$, $(6.8\pm0.9)\times 10^{-3}$, and
$(1.4\pm0.1)\times 10^{-2}$ counts per second for the first, second, and third
{\it Chandra} observation, on MJD 53371.88 (Jan.~1, 2005), 53383.99 (Jan.~13,
2005), and 53407.57 (Feb.~6, 2005), respectively. The count rate for the third
observation is $2.0\pm0.4$ times as high as that of the average of the first
two observations. The unabsorbed 0.5--10 keV source flux for the best--fit
power--law model to the source spectrum was $(7.9\pm2.5)\times 10^{-14}$ erg
cm$^{-2}$ s$^{-1}$, $(7.3\pm2.0)\times 10^{-14}$ erg cm$^{-2}$ s$^{-1}$, and
$(1.17\pm0.22)\times 10^{-13}$ erg cm$^{-2}$ s$^{-1}$ for the first, second,
and third {\it Chandra} observation, respectively. We find that this source
flux is consistent with that found by ROSAT [$\approx (5.4\pm2.4)\times
10^{-14}$ erg cm$^{-2}$ s$^{-1}$]. Under the assumption that the interstellar
extinction, N$_H$, does not vary between the observations, we find that the
blackbody temperature during the second {\it Chandra} observation is
significantly higher than that during the first and third observation.
Furthermore, the effective temperature of the neutron star derived from fitting
an absorbed blackbody or neutron star atmosphere model to the data is rather
high in comparison with many other neutron star soft X--ray transients in
quiescence, even during the first and third observation. If we assume that the
source quiescent luminosity is similar to that measured for two other accretion
powered millisecond pulsars in quiescence, the distance to IGR~J00291+5934 is
2.6--3.6 kpc.

\end{abstract}


\section{Introduction} 

The evolutionary link between the millisecond radio pulsars and the low--mass X--ray binaries was
established by the detection of the first accretion--powered millisecond X--ray pulsar SAX~J1808.4--3658
(\citealt{1998Natur.394..344W}; \citealt{1998Natur.394..346C}). Over the last years the number of known
accretion--powered millisecond X--ray pulsars has steadily increased. Recently, the discovery of the
sixth member of this class was announced. The source IGR~J00291+5934 was discovered by INTEGRAL
(\citealt{2004ATel..352....1E}; \citealt{2005A&A...432L..13S}) on Dec.~2, 2004 (MJD~53341). Using {\it
Rossi} X--ray Timing Explorer (RXTE) follow--up observations, \citet{2004ATel..353....1M} discovered
pulsations at a frequency of 598.88~Hz. The pulsar is in a 147.4~min.~orbit with a low--mass companion
star (\citealt{2004ATel..360....1M}; \citealt{2005ApJ...622L..45G}). A previously undetected source was
found at a magnitude of $R \approx 17.4$ within the INTEGRAL error circle in an $R$--band image obtained
with the Robotic Palomar 60--inch telescope (\citealt{2004ATel..354....1F}). Spectra obtained with the
4.2~m William Herschel Telescope of this variable source showed H$\alpha$ and He~II emission lines,
securing the identification of this variable star as the counterpart of IGR~J00291+5934
(\citealt{2004ATel..356....1R}; see also \citealt{2004ATel..366....1F}). Finally, a variable radio and 
near infra--red source was found at the position of the optical counterpart
(\citealt{2004ATel..355....1P}; \citealt{2004ATel..361....1F}; \citealt{2004ATel..363....1S}).

During outburst the X--ray, optical, and radio properties of the accretion--powered X--ray pulsars are very
similar to those of non--pulsating neutron star soft X--ray transients (SXTs;
e.g.~\citealt{1999ApJ...514..939W} for an X--ray account). In quiescence, the accretion powered X--ray
pulsars observed so far with {\it Chandra} and/or XMM--{\it Newton} have lower luminosities than those of
many non--pulsating neutron star SXTs in quiescence. Furthermore, in those quiescent systems where it has
been possible to detect sufficient photons in X--rays  to allow for a spectral decomposition, the X--ray
spectrum is significantly harder than for the non--pulsating neutron star SXTs 4U~1608--52, Aql~X--1,
SAX~J1748.8--2021 in the Globular Cluster NGC~6440, and XTE~J1709-267 in quiescence
(\citealt{1996PASJ...48..257A}; \citealt{2002ApJ...577..346R}; \citealt{2002ApJ...575L..15C};
\citealt{2003MNRAS.341..823J}; \citealt{2005ApJ...619..492W}; \citealt{2001ApJ...563L..41I};
\citealt{2005ApJ...620..922C}; see also \citealt{2004MNRAS.354..666J} and references therein). However,
there are systems that have intermediate spectral properties and luminosities
(e.g.~\citealt{2004MNRAS.349...94J}). Out of these sources with intermediate properties Cen~X--4
(\citealt{2001ApJ...551..921R}), GRS~1741.9--2853 (\citealt{2003ApJ...598..474M}), and SAX~J1810.8--2609
(\citealt{2004MNRAS.349...94J}) have not been observed in outburst with the RXTE Proportional Counter
Array. For that reason it is unclear whether pulsations were present in outburst or not.

The spectrum of several neutron star SXTs in quiescence is found to be well--fit by a neutron star
atmosphere (NSA) model sometimes supplemented with a power--law component. Especially in sources with a
quiescent luminosity near $10^{33}$ erg s$^{-1}$ the spectrum is dominated by a strong thermal component
(\citealt{2004MNRAS.354..666J}). \citet{1998ApJ...504L..95B} and \citet{2001ApJ...548L.175C} showed that
the neutron star core is heated by pycnonuclear reactions during/immediately after accretion episodes. In
quiescence the heated neutron star core cools in X--rays. For a given core temperature, from the assumed
thermal properties of the neutron star envelope, the cooling core spectral energy distribution is
determined by the neutron star atmosphere. In theory, an NSA model fit to the quiescent thermal X--ray
component provides the means to measure the mass and radius of the neutron star and hence constrain the
equation of state of matter at supranuclear densities. In practice, numbers typical for a canonical neutron
star were found (e.g.~\citealt{2003ApJ...598..501H}), rendering support for this interpretation. However,
there is an ongoing debate about whether the temperature of the thermal component is varying or not
(cf.~\citealt{2002ApJ...577..346R}; \citealt{2003ApJ...597..474C}; \citealt{2004ApJ...601..474C}).
Differences in temperatures obtained within one quiescent epoch may be explained by changes in the neutron
star atmosphere due to ongoing low--level accretion (\citealt{2002ApJ...574..920B}). Similarly, differences
in the observed neutron star temperature obtained before and after outburst  activity can be explained by
differences in the amount of residual unburned light elements in the neutron star atmosphere
(\citealt{2002ApJ...574..920B}). Large changes on short timescales would render it unlikely that the
soft/thermal component is due to cooling of the neutron star, limiting the applicability of the NSA model
fit. Finally, there are currently two sources known which returned to quiescence after an accretion epoch
lasting several years (i.e.~KS~1731--260 and MXB~1659--298). In those cases it is thought that the
observed  thermal spectral component is caused by the cooling of the neutron star {\it crust}
(\citealt{2002ApJ...573L..45W}; \citealt{2002ApJ...580..413R}). The observed changes in the X--ray spectral
properties are also ascribed to the cooling down of the neutron star crust (\citealt{2004wijninpress}). 

In this Paper we present the detailed analysis of three {\it Chandra} observations of the
accretion powered X--ray pulsar IGR~J00291+5934 obtained during the late stages of the 2004
outburst.

\section{Observations, analysis, and results} 

We observed IGR~J00291+5934 three times after the 2004 discovery outburst with the
back--illuminated S3 CCD--chip of the Advanced CCD Imaging Spectrometer (ACIS) detector on board
the {\it Chandra} satellite. The observations started on MJD~53371.88 (Jan.~1, 2005; {\it Chandra}
observation ID 6179), MJD~53383.99 (Jan.~13, 2005; observation ID 6180), and MJD~53407.57 (Feb.~6,
2005; observation ID 6181; all times are in UTC). The net, on--source exposure times were 4.7,
9.0, and 12.9 ksec, respectively. We limited the readout area of the S3--chip to 1/8th of its
original size yielding a smaller exposure time per CCD frame in order to avoid pile--up.

After the data were processed by the {\it Chandra} X--ray Center (ASCDS version 7.5.0), we analysed them
using the {\it CIAO 3.2.1} software developed by the Chandra X--ray Center. We searched the data for
background flares but none were found, hence we used all data in our analysis. We extracted data from a
circular area with a radius of three pixels ($\sim1.5"$) centred on the best--fit source position for
the first two observations. This area encloses 90--95 per cent of the energy since, as we will show
below, the source spectrum is not very hard. The reason why we use this small area is to exclude as much
as possible background photons. In such a small area one expects 1$\pm$1 background photon for the first
observation. During the third observation (MJD~53407.57) the count rate is higher than in the previous
two observation (see below), which led us to use a six pixel ($\sim3"$) radius circular area centred on
the source position for the source extraction. 

In the three observations we detect respectively 36, 63, and 190 photons from a position consistent with
that found in the optical by \citet{2004ATel..354....1F} and in X--rays by \citet{2004ATel..369....1N}.
This yields 0.3--10 keV count rates of $(7.2\pm1.2)\times 10^{-3}$, $(6.8\pm0.9)\times 10^{-3}$, and
$(1.4\pm0.1)\times 10^{-2}$ counts per second. These background subtracted (see below) count rates are so
low that pile--up is not a concern. The count rate for the third observation is $2.0\pm0.4$ times as high
as that of the average of the first two observations.  Due to the low number of source counts the Cash
statistic was used for the spectral analysis of the first two observations (\citealt{1979ApJ...228..939C}).
This means that the background should not be subtracted when fitting the spectrum in order to maintain the
counting statistics (although with a background as low as that in our case, this should not make much
difference). For the extraction of the background spectrum in the third observation, we used an annulus
with radius of 5--15". We rebinned the source spectrum of the third observation such that each bin contains
at least 10 counts. We fitted the spectra in the 0.3--10 keV range using {\sc XSPEC}
(\citealt{1996adass...5...17A}) version 11.3.1. Since 10 counts per bin is still low for the
use of $\chi^2$ statistics in the spectral fits of the third observation, we compared the results using the
$\chi^2$ statistics with those obtained using the Cash statistics; the results are the same within the
errors.

\begin{figure} \includegraphics[width=8cm]{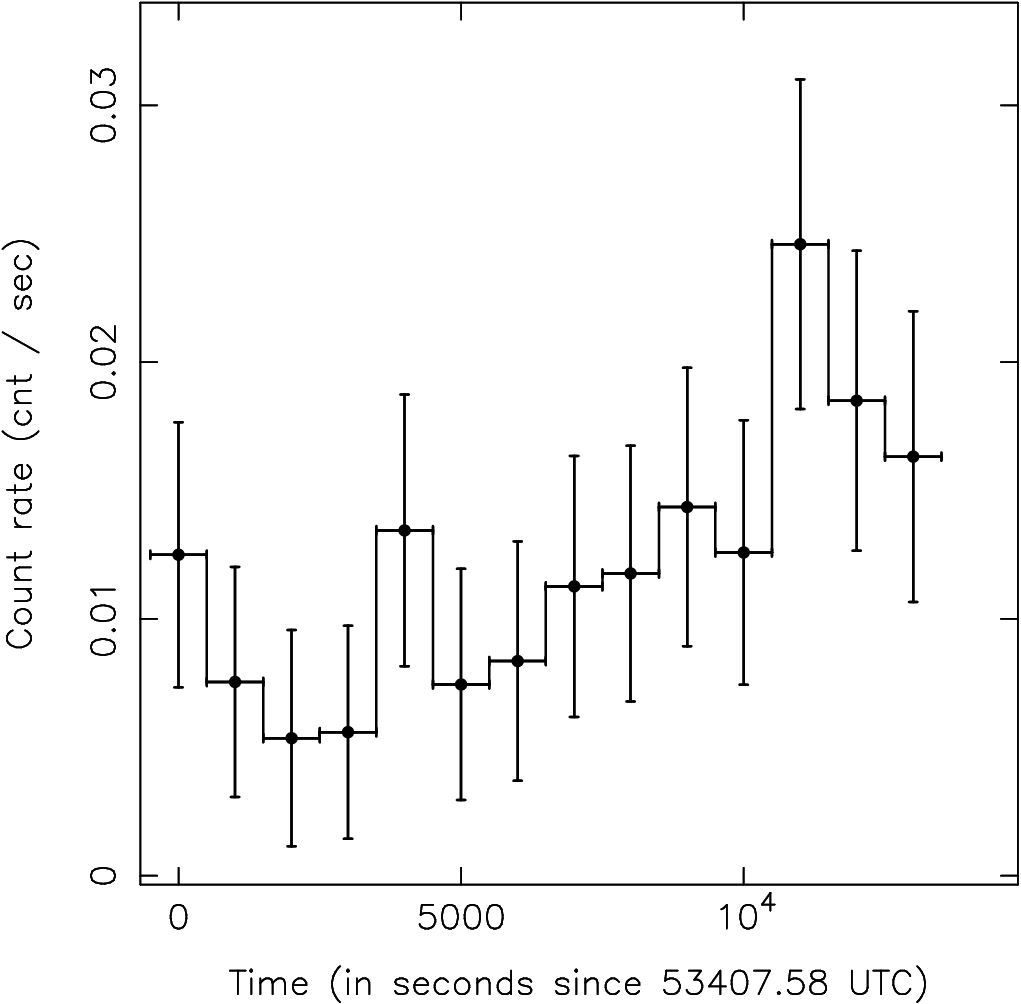} \caption{The 0.3--10 keV background subtracted
lightcurve for the accretion--powered millisecond X--ray pulsar IGR~J00291+5934 detected with the {\it
Chandra} satellite on MJD~53407.57. The data were binned in 1000~second intervals. Variability in the
count rate as a function of time within the observation is clearly detected.}  \label{lightc}  \end{figure}

For each observation separately we searched for variability in the lightcurve. Only in the last
observation (ID 6181) is there evidence that the source count rate was higher towards the end of the
observation than near the start of the observation (see Figure~\ref{lightc}). A Kolmogorov--Smirnov test
(see \citealt{prteve1992}) testing the hypothesis that the count rate of the source is constant showed
that the probability that the count rate is constant is 0.36, 0.36, and 8.4$\times 10^{-5}$ for
observation with ID 6179, 6180, 6181, respectively (i.e.~the hypothesis is observed to be disproven at the
4.1$\sigma$ level for the third observation). We searched for spectral variability within an observation
by making hardness ratios for the first and second half of the total number of photons detected during
each observation and by investigating the photon energy as a function of photon arrival time. The hardness
is defined as the count rate ratio between the 0.3--1.5 keV band and the 1.5--10 keV band. There is no
evidence for significant changes in the hardness ratio during any of the observations. To check whether
the hardness ratio changes between the observations we calculated the hardness ratio from all photons
detected during one observation. The hardness ratio during the first, second and third observation is
2.8$\pm$1.1, 1.3$\pm$0.3, and 2.3$\pm$0.4, respectively. The hardness ratio of the third observation is
larger  than that of the second observation, albeit at the 2~$\sigma$ level.

In order to check whether the spectrum changed between the observations we fitted the spectra for each of
the observations separately. We fit the spectra using several functions that are often used for quiescent
neutron stars: a power--law, an NSA, and a black--body model, all of these were modified by the effects of
interstellar absorption. In these fits the value of the interstellar absorption, ${\rm N_H}$, was fixed to
2.8$\times 10^{21}{\rm cm^{-2}}$ similar to the value found by \citet{2004ATel..369....1N}. When we left
the ${\rm N_H}$ as a free parameter in the spectral fits for the third observation the ${\rm N_H}$ was
consistent with 2.8$\times 10^{21}{\rm cm^{-2}}$. As an indication of the influence that fixing the N$_H$
has on the fit parameters we include the results of a blackbody, a power--law, and a blackbody fit to the
first, second, and third observation, respectively, in Table~1.  For the NSA model the neutron star radius
and mass  are fixed to 10 km and 1.4 M$_\odot$, respectively. Furthermore, we assumed that the neutron star
magnetic field is so low that it is unimportant for the NSA modelling, leaving only the normalisation
(i.e.~one divided by the source distance squared) and the effective temperature as free parameters in this
model. The results of these fits are shown in Table~1. 

The likelihood ratio method, which is the basis of the Cash statistic, does not provide a direct test to
the goodness--of--fit which would help one to distinguish between different models
(e.g.~\citealt{1992drea.book.....B}). In order to obtain a handle on the goodness--of--fit we used Monte
Carlo simulations. They consist of simulating 10$^4$ counts spectra. Each spectrum is drawn from a parent
distribution with Poisson statistics. The parent distribution is allowed to vary according to the
covariances of the best--fit model parameters. The C fit--statistic for each of the Monte Carlo simulations
is compared with that of the best--fit model. After the Monte Carlo simulations have been performed the
fraction of simulations that gave a lower fit statistic than that of the best--fit model is given. This
percentage is called the {\sl goodness}. If the model provides a good description of the data the goodness
should be close to 50 per cent. Both a very low goodness and a very high goodness indicate that the model
does not represent the data accurately. As can be seen in Table~1, for many of the single component models
for which N$_H$ was held fixed at 2.8$\times 10^{21}{\rm cm^{-2}}$ the goodness is not close to the nominal
50 per cent. However, in none of the case we can exclude the model at high significance based on the
goodness.

\begin{figure} 
\includegraphics[width=6.5cm,angle=-90]{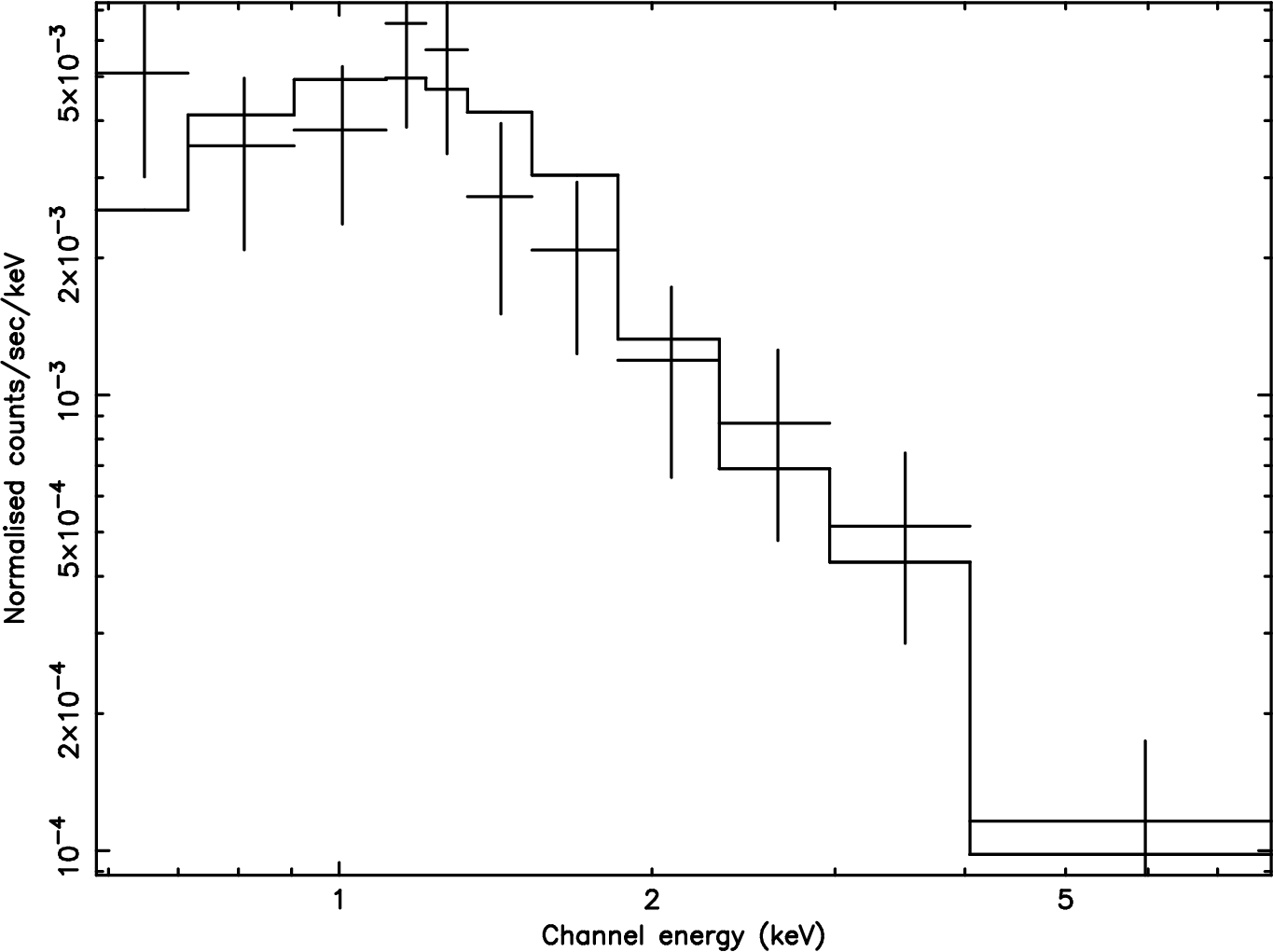} 
\quad
\includegraphics[width=6.5cm,angle=-90]{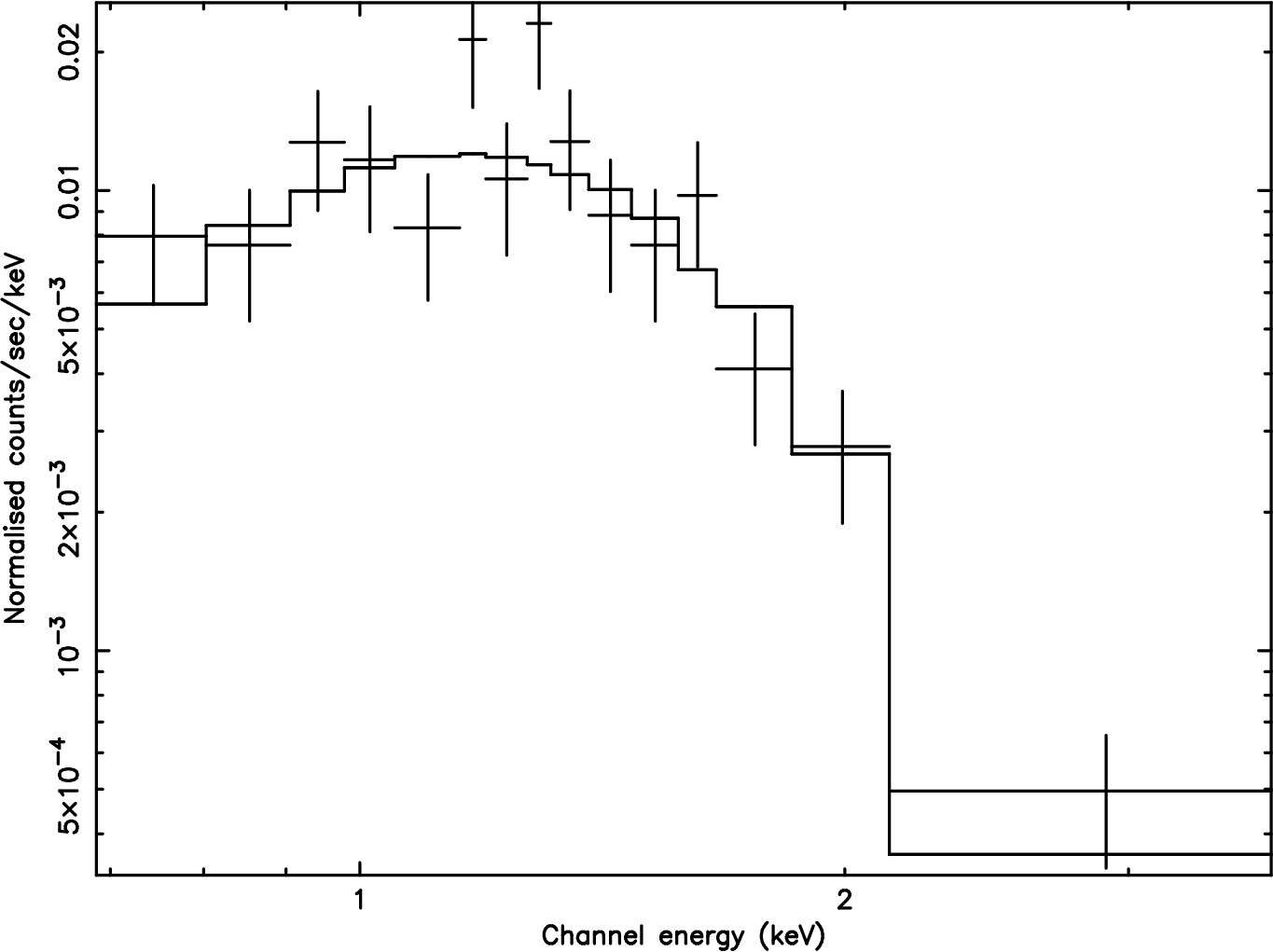}

\caption{{\it Top panel:} The quiescent spectrum of IGR~J00291+5934 detected with the {\it Chandra}
satellite on MJD~53383.99 (Jan.~13, 2005). The best fitting power--law model is over plotted. The fit
was done on the unbinned data using the C--statistic, but for clarity the data points have been rebinned
in the plot. {\it Bottom panel:} The quiescent spectrum of IGR~J00291+5934 detected with the {\it
Chandra} satellite on MJD~53407.57 (Feb.~6, 2005). The best fitting blackbody model is over plotted. In
both cases the N$_H$ was fixed to 2.8$\times 10^{21}$ cm$^{-2}$.}  \label{spectra}   \end{figure}

The spectral fits suggest that the spectrum of the source is harder during the second observation than
during the first and third observation, as can be seen from a comparison between the best--fit power--law
index or the blackbody/NSA temperature of the observations. Besides the single component models we tried
fitting a model comprised of a blackbody and a power law modified by the effects of interstellar absorption
to the data in order to assess the contribution of a power law to the spectrum. However, due to the low
number of detected photons during the first and second observation the fit parameters were unconstrained. A
fit using such a fit function resulted in meaningful constraints only for the third observation (and only
when we kept the power--law index fixed to 2). The results of this fit are also listed in Table~1. We also
tried to fit an NSA plus power--law model to the spectrum of the third observation. However, with the
power--law index fixed to a value of 2 the power--law did not contribute significantly to the fit (the
normalisation was a factor of 10$^3$ less than that found for the blackbody plus power law fit).  In
Figure~\ref{spectra} we have plotted the spectrum showing a power--law fit to the dataset of the second
{\it Chandra} observation and a blackbody fit to the third {\it Chandra} observation.

\begin{figure} \includegraphics[width=8cm]{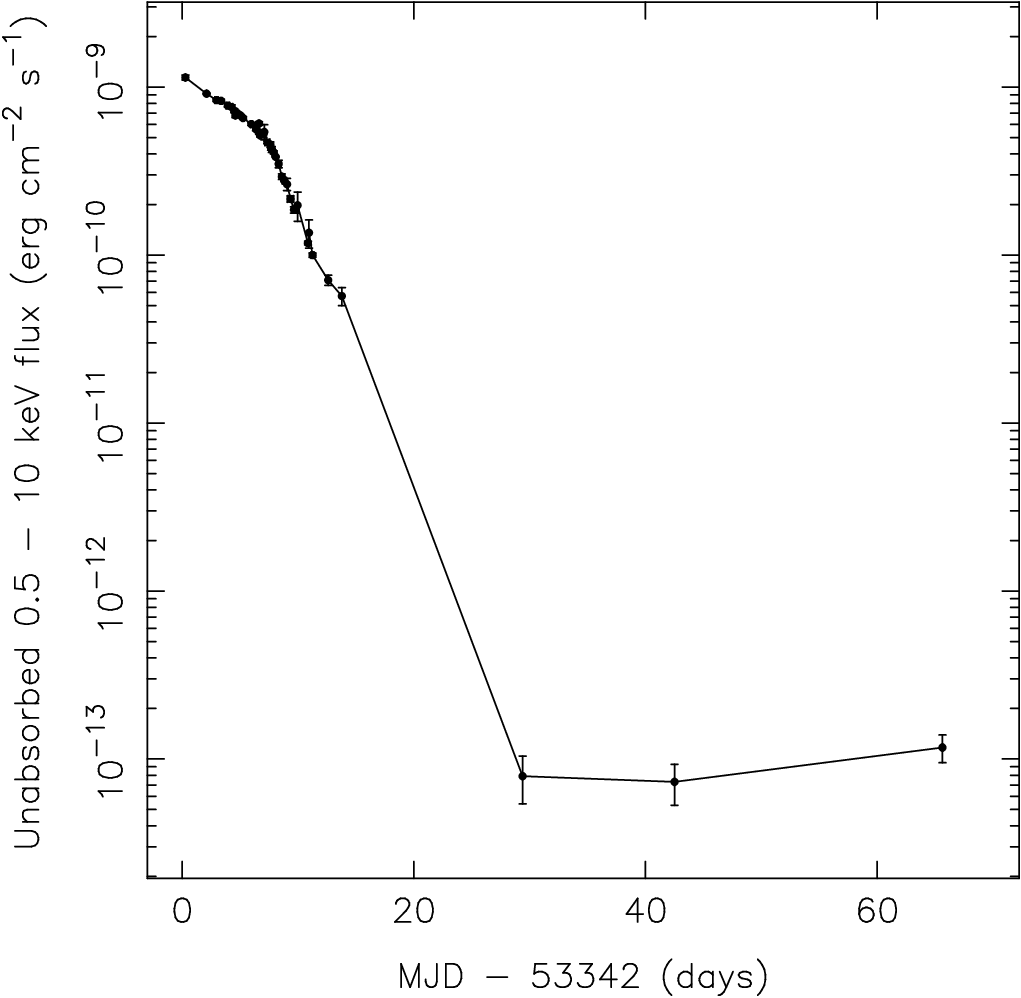} \caption{The 0.5--10 keV unabsorbed
2004/2005 outburst lightcurve for the accretion--powered millisecond X--ray pulsar IGR~J00291+5934 as
detected with the RXTE and {\it Chandra} satellites. The RXTE 2.5--25 keV observed flux (Galloway et
al.~2005) is converted to an unabsorbed 0.5--10 keV flux using a power--law spectrum with index 2 and an
interstellar absorption of ${\rm N_H=2.8\times 10^{21} cm^{-2}}$.}  \label{lc}  \end{figure}

The unabsorbed 0.5--10 keV source flux for the first, second, and third {\it Chandra} observation using
the best--fit power--law model is $(7.9\pm2.5)\times 10^{-14}$ erg cm$^{-2}$ s$^{-1}$,
$(7.3\pm2.0)\times 10^{-14}$ erg cm$^{-2}$ s$^{-1}$, and $(1.17\pm0.22)\times 10^{-13}$ erg cm$^{-2}$
s$^{-1}$, respectively (see Figure~\ref{lc}; the plotted uncertainty on these fluxes is at the 68 per
cent confidence level). We also listed in Table~1 the unabsorbed 0.5--10 keV flux for each model for
each observation. In between brackets we give the upper limit to the fractional contribution to the
unabsorbed 0.5--10 keV flux for a power--law model in case of a blackbody or NSA model fit. We also give
the upper limit to the fractional contribution for a blackbody and NSA model in case of a power--law
fit.

Finally, we extracted an 18 ksec.~archival Positional Sensitive Proportional Counter (PSPC) {\it ROSAT}
observation of the Cataclysmic Variable RX~J0028.8+5917 also known as DQ~Her obtained over the period
MJD 48830--48839 (July~26--August 4, 1992). Besides DQ~Her, a source is detected at a position
consistent within the {\it ROSAT} PSPC positional accuracy with that of IGR~J00291+5934. There are 44
counts in a circular region with radius of 40" in the 0.1--2 keV energy range. However, from a
background determination of a circular region 100" away, we find that 43 per cent of those counts can be
attributed to the background. This yields a background subtracted source count rate of 
$(1.4\pm0.4)\times 10^{-3}$ counts per second in the PSPC instrument. If we use a power law with index
2.5 and fix the ${\rm N_H}$ to 2.8$\times 10^{21}{\rm cm^{-2}}$ we get an unabsorbed 0.5--10 keV source
flux of $\approx (5.4\pm2.4)\times 10^{-14}$ erg cm$^{-2}$ s$^{-1}$ at the time of the {\it ROSAT}
observation.

\begin{sidewaystable*}  {\normalsize  Table 1: Best fit parameters of the quiescent spectrum of
IGR~J00291+5934. NSA stands for neutron star atmosphere, BB refers to blackbody, and PL to power law.  All
quoted errors are at the 90 per cent confidence level. The value in between brackets in the flux column
denotes the (95 per cent upper limit to the) fractional contribution of a PL in case of a BB or NSA model
fit or a BB/NSA in case of a PL model fit. In case of an upper limit to the fractional PL contribution the
PL index was fixed at 2. In case of an upper limit to the fractional BB/NSA contribution the BB/NSA
temperature was fixed at 0.2 keV/10$^{6}$ K. For the first two observations the goodness--of--fit is
expressed via the goodness parameter, whereas for the third observation the goodness--of--fit is given by
means of the reduced $\chi^2$ for a certain degrees of freedom (d.o.f.).}

{\tiny


\label{spec}
\begin{tabular}{lccccccclc}
\hline
Obs ID& MJD & Model & N$_H~\times10^{21}$& BB radius & NSA norm. & PL norm. $\times10^{-5}$& Temp. / PL index & Unabs. 0.5--10 keV flux & Goodness / ${\rm \chi^2_{red}}$ \\ 
& (UTC) &  & cm$^{-2}$&  ${\rm (\frac{d}{10\,kpc})^2}$ km & ${\rm (\frac{10\,kpc}{d})^2}$ & phot.
keV$^{-1}$~cm$^{-2}$~s$^{-1}$ & BB (keV), NSA (log K) & (erg cm$^{-2}$ s$^{-1}$) & per cent / d.o.f.\\
\hline
\hline
6179 & 53371.88 &  BB  & 2.8$^a$ &  1.1$^{+1.9}_{-0.7}$ & --- & --- &  0.27$\pm$0.05 & 5.2$\times 10^{-14}$~($<43$\%$^b$) & 6\% \\
6179 & 53371.88 &  NSA & 2.8$^a$  & --- & 0.21$^{+1.01}_{-0.17}$ & --- & 6.25$\pm$0.15 & 5.5$\times 10^{-14}$~($<39$\%$^b$) & 34\% \\
6179 & 53371.88 &  PL  & 2.8$^a$  & --- & --- & $2.6\pm0.8$ & 3.2$\pm$0.7 & 7.9$\times 10^{-14}$~($<23/26$\%$^b$) & 3\% \\
6179 & 53371.88 &  BB  & 0.8$_{-0.8}^{+3}$ &  0.36$^{+3.6}_{-0.26}$ & --- & --- &  0.32$\pm$0.10 & 3.5$\times 10^{-14}$~($<47$\%$^b$) & 31\% \\
\hline
6180 & 53383.99 &  BB     & 2.8$^a$  & $(5.8_{-2.7}^{+4.6})\times 10^{-2}$& --- & --- & 0.54$\pm$0.09 & 5.3$\times 10^{-14}$~($<68$\%$^b$) & 99\% \\
6180 & 53383.99 &  NSA    & 2.8$^a$  & --- & $(0.40_{-0.27}^{+0.46})\times 10^{-2}$ & --- & 6.65$\pm0.10$& 5.7$\times 10^{-14}$~($<53$\%$^b$) & 88\% \\
6180 & 53383.99 &  PL     & 2.8$^a$  & --- & --- & $1.8\pm0.5$ & 2.3$\pm0.4$& 7.3$\times 10^{-14}$~($<15/18$\%$^b$) & 14\% \\
6180 & 53383.99 &  PL     & 0.7$_{-0.7}^{+2.4}$  & --- & --- & $1.0_{-0.4}^{+1.0}$ & 1.7$\pm0.6$& 6.3$\times 10^{-14}$~($<10/10$\%$^b$) & 24\% \\

\hline
6181 & 53407.57 &  BB     & 2.8$^a$  & 1.0$\pm0.5$& --- & --- & 0.31$\pm$0.03 & 8.3$\times 10^{-14}$~($<27$\%$^b$) & 1.0/13 \\
6181 & 53407.57 &  NSA    & 2.8$^a$  & --- & $0.15_{-0.1}^{+0.14}$ & --- & 6.33$^{+0.11}_{-0.06}$ & 9.0$\times 10^{-14}$~($<25$\%$^b$) &  0.96/13 \\
6181 & 53407.57 &  PL     & 2.8$^a$  & --- & --- & $4.1\pm0.6$&  2.9$\pm$0.3& 1.3$\times 10^{-13}$~($<12/12$\%$ ^b$) & 1.4/13 \\
6181 & 53407.57 &  BB+PL  & 2.8$^a$  & 1.16$\pm0.57$ & --- & $0.7_{-0.7}^{+15}$& 0.28$\pm0.05$ / 2$^a$ & 1.0$\times 10^{-13}$~(67+33)$\%^b$ & 1.0/12 \\
6181 & 53407.57 &  BB     & 1.3$_{-1.3}^{+2.7}$&0.48$_{-0.3}^{+1.5}$& --- & --- & 0.34$\pm$0.07 & 6.5$\times 10^{-14}$~($<25$\%$^b$) & 1.0/12 \\

\end{tabular}

{\footnotesize$^a$ Parameter fixed.}\\
{\footnotesize$^b$ Percentage contributed to the total flux / upper limit to the fractional BB/NSA or PL contribution to the total flux. }\\
}
\end{sidewaystable*}

\section{Discussion}

We have presented the analysis of three {\it Chandra} observations of the accretion--powered millisecond
X--ray pulsar IGR~J00291+5934. The observations were obtained right after the end of the 2004 discovery
outburst. The count rate for the third observation is $2.0\pm0.4$ times higher than that of the average
of the first two observations. This shows that even in quiescence, there is evidence for variability (at
the $\sim$5~$\sigma$ level). However, owing to spectral variability between the three observations, the
unabsorbed 0.5--10 keV source flux of the {\it Chandra} observations did not change significantly. We
also detected the source in quiescence in a {\it ROSAT} observation obtained in 1992. The flux that we
measured in the {\it Chandra} observations does not differ significantly from that in the 1992 {\it
ROSAT} observation. From this we derive that the source had reached quiescence already during our first
{\it Chandra} observation obtained $\sim$31 days after the discovery of the source
(\citealt{2005A&A...432L..13S}).

Under the assumption that the interstellar extinction, N$_H$, does not vary between the observations, we
find from a blackbody/NSA fit to the spectrum of the second observation that the blackbody/NSA temperature
is significantly higher than that during the first and third observation. Furthermore, a single model fit
using either an NSA or a blackbody model with the N$_H$ fixed to the value derived during outburst
(N$_H=2.8\times 10^{21}$ cm$^{-2}$; \citealt{2004ATel..369....1N}) yields a rather high temperature when
compared with other neutron star soft X--ray transients in quiescence even for the first and third
observation (cf.~T${\rm _{BlackBody}}$=0.27$\pm$0.05, 0.31$\pm$0.03 keV/ log~${\rm (T_{NSA})}=6.25\pm0.15$,
6.33$^{+0.11}_{-0.06}$ for the first and third observation of IGR~J00291+5934, respectively. Whereas
log~${\rm T_{NSA}}=5.84-6.14$ for Aql~X--1 [range due to variability], \citealt{2002ApJ...577..346R};
T${\rm _{BlackBody}}$=0.186$\pm0.005$ keV / log~${\rm (T_{NSA})}=5.99\pm0.02$ for Cen~X--4,
\citealt{2004ApJ...601..474C}; T${\rm _{BlackBody}}$=0.14$\pm$0.02 keV for SAX~J1810.8--2609,
\citealt{2004MNRAS.349...94J}; T${\rm _{BlackBody}}$=0.24$\pm$0.02 keV for XTE~J1709--267,
\citealt{2004MNRAS.354..666J}; log~${\rm (T_{NSA})}=6.02-6.24$ for SAX~J1748.8--2021 in the Globular
Cluster NGC~6440 [range due to variability], \citealt{2005ApJ...620..922C}).  Even when we leave the N$_H$
as a free parameter the best--fitting blackbody temperature is 0.32$\pm$0.06 keV for the third observation.
Such a high temperature can be explained if the source spectrum contains a harder component besides a
single (absorbed) thermal component. For the last and longest observation, the signal--to--noise ratio is
high enough to test this by fitting the data with a model consisting of a blackbody and a power law,
modified by the effects of interstellar absorption. This provides a good fit with a slightly lower
blackbody temperature of 0.28$\pm$0.05 keV when we fix the power--law index to 2. In this case the power
law contributes 33 per cent to the source luminosity.

\citet{2002ApJ...577..346R} and \citet{2004ApJ...601..474C} showed that the quiescent spectra and flux
of Aql~X--1 and Cen~X--4 also vary during and between observations. However, in their interpretation
these authors differ. \citet{2003ApJ...597..474C} showed that with the current data for Aql~X--1, it is
not possible to distinguish between intrinsic source variability in the soft blackbody component and
correlated variability in the interstellar column density and the power--law photon index. Here we show,
for the first time, that besides in these two canonical neutron star soft X--ray transients, spectral
and count rate variability also occurs in the quiescent X--ray emission of an accretion powered
millisecond X--ray pulsar. Significant variability in the lightcurve was also found during the last
observation. At the 95 per cent confidence level there is evidence for variability during an observation
of the accretion--powered millisecond X--ray pulsar XTE~J0929--314 (\citealt{2005ApJ...619..492W}). As
explained in the Introduction, the variability observed in the quiescent flux of the neutron star SXTs
KS~1731--260 and MXB~1659--29 (\citealt{2002ApJ...573L..45W}; \citealt{2004wijninpress}) likely has a
different origin than the variability observed here. Similarly, the variability of SAX~J1748.8--2021 in
the Globular Cluster NGC~6440 in quiescence (\citealt{2005ApJ...620..922C}) can be caused by the fact
that in between the two quiescent observations the source experienced an outburst. The quiescent flux of
GRS~1741.9--2853 was found to vary by more than a factor of 5 (\citealt{2003ApJ...598..474M}). However,
this source has, so far, never been observed in outburst with RXTE and hence it is unclear whether this
source contains an accretion--powered millisecond X--ray pulsar.

The ratio between the peak and lowest quiescent unabsorbed 0.5--10 keV X--ray flux is $\sim2\times
10^{4}$. This was found after correcting the peak flux given in \citet{2005ApJ...622L..45G} for the different 
band--pass assuming the peak--flux spectrum is similar to the spectrum found by
\citet{2004ATel..369....1N}. This ratio is low for a soft X--ray transient. Unfortunately, the distance
to IGR~J00291+5934 is not well constrained, so in principle we cannot say whether the quiescent
luminosity is high or the outburst luminosity is low (or both). \citet{2005ApJ...622L..45G} mention a minimum
distance of 4 kpc but as those authors also state, the uncertainty in this estimate is large. If we
assume that IGR~J00291+5934 has a 0.5--10 keV quiescent luminosity of $\approx5-10\times 10^{31}$ erg
s$^{-1}$, similar to that of the accretion powered milliseconds pulsars SAX~J1808.4-3658 and
XTE~J0929--314 (\citealt{2002ApJ...575L..15C}; \citealt{2005ApJ...619..492W}), the distance to
IGR~J00291+5934 would be $\approx$2.6--3.6 kpc. In light of the uncertainty of this estimate the values
derived by us and Galloway et al.~(2005) are close together. However, an NSA model does not provide a
satisfactory fit to the spectrum of our third {\it Chandra} observation if we fix the model
normalisation to that obtained for a distance of 3.5 kpc ($\chi^2_{red}=3.7$ for 14 degrees of freedom).
If we let the N$_H$ free the $\chi^2_{red}=1.5$ for 13 degrees of freedom with
N$_H=0.65_{-0.08}^{+0.28} \times 10^{21}$ cm$^{-2}$ for a NSA temperature of 94$\pm4$ eV.

Since this source is relatively bright in X--rays in quiescence, a future, longer XMM--{\it Newton} or {\it
Chandra} observation should provide better constraints on the spectral parameters and test for the time
scale of variability.

\section*{Acknowledgments}  

\noindent PGJ would like to thank the {\it Chandra} Director, Harvey Tananbaum, for the allocation
of Directors Discretionary Time and the {\it Chandra} support team for help with preparing the
observations and the fast prossessing of the data. The authors would like to thank the referee for
useful comments and suggestions which helped improve the paper and Mariano M\'endez for reading and
commenting on an earlier version of the manuscript. Support for this work was provided by NASA
through Chandra Postdoctoral Fellowship grant number PF3--40027 awarded by the Chandra X--ray
Center, which is operated by the Smithsonian Astrophysical Observatory for NASA under contract
NAS8--39073. DS acknowledges a Smithsonian Astrophysical Observatory Clay Fellowship. PGJ further
acknowledges support from NASA grant GO4-5033X Fund number 16617404.


\begin{thebibliography}{39}
\expandafter\ifx\csname natexlab\endcsname\relax\def\natexlab#1{#1}\fi

\bibitem[{{Arnaud}(1996)}]{1996adass...5...17A}
{Arnaud}, K.~A., 1996, in ASP Conf. Ser. 101: Astronomical Data Analysis
  Software and Systems V, vol.~5, p.~17

\bibitem[{{Asai} et~al.(1996){Asai}, {Dotani}, {Mitsuda}, {Hoshi}, {Vaughan},
  {Tanaka}, \& {Inoue}}]{1996PASJ...48..257A}
{Asai}, K., {Dotani}, T., {Mitsuda}, K., {Hoshi}, R., {Vaughan}, B., {Tanaka},
  Y., {Inoue}, H., 1996, \pasj, 48, 257

\bibitem[{{Bevington} \& {Robinson}(1992)}]{1992drea.book.....B}
{Bevington}, P.~R., {Robinson}, D.~K., 1992, {Data reduction and error analysis
  for the physical sciences}, New York: McGraw-Hill, |c1992, 2nd ed.

\bibitem[{{Brown} et~al.(1998){Brown}, {Bildsten}, \&
  {Rutledge}}]{1998ApJ...504L..95B}
{Brown}, E.~F., {Bildsten}, L., {Rutledge}, R.~E., 1998, \apjl, 504, L95

\bibitem[{{Brown} et~al.(2002){Brown}, {Bildsten}, \&
  {Chang}}]{2002ApJ...574..920B}
{Brown}, E.~F., {Bildsten}, L., {Chang}, P., 2002, \apj, 574, 920

\bibitem[{{Cackett} et~al.(2005)}]{2005ApJ...620..922C}
{Cackett}, E.~M., et~al., 2005, \apj, 620, 922

\bibitem[{{Campana} \& {Stella}(2003)}]{2003ApJ...597..474C}
{Campana}, S., {Stella}, L., 2003, \apj, 597, 474

\bibitem[{{Campana} et~al.(2004){Campana}, {Israel}, {Stella}, {Gastaldello},
  \& {Mereghetti}}]{2004ApJ...601..474C}
{Campana}, S., {Israel}, G.~L., {Stella}, L., {Gastaldello}, F., {Mereghetti},
  S., 2004, \apj, 601, 474

\bibitem[{{Campana} et~al.(2002)}]{2002ApJ...575L..15C}
{Campana}, S., et~al., 2002, \apjl, 575, L15

\bibitem[{{Cash}(1979)}]{1979ApJ...228..939C}
{Cash}, W., 1979, \apj, 228, 939

\bibitem[{{Chakrabarty} \& {Morgan}(1998)}]{1998Natur.394..346C}
{Chakrabarty}, D., {Morgan}, E.~H., 1998, \nat, 394, 346

\bibitem[{{Colpi} et~al.(2001){Colpi}, {Geppert}, {Page}, \&
  {Possenti}}]{2001ApJ...548L.175C}
{Colpi}, M., {Geppert}, U., {Page}, D., {Possenti}, A., 2001, \apjl, 548, L175

\bibitem[{{Eckert} et~al.(2004){Eckert}, {Walter}, {Kretschmar}, {Mas-Hesse},
  {Palumbo}, {Roques}, {Ubertini}, \& {Winkler}}]{2004ATel..352....1E}
{Eckert}, D., {Walter}, R., {Kretschmar}, P., {Mas-Hesse}, M., {Palumbo},
  G.~G.~C., {Roques}, J.-P., {Ubertini}, P., {Winkler}, C., 2004, The
  Astronomer's Telegram, 352, 1

\bibitem[{{Fender} et~al.(2004){Fender}, {De Bruyn}, {Pooley}, \&
  {Stappers}}]{2004ATel..361....1F}
{Fender}, R., {De Bruyn}, G., {Pooley}, G., {Stappers}, B., 2004, The
  Astronomer's Telegram, 361, 1

\bibitem[{{Filippenko} et~al.(2004){Filippenko}, {Foley}, \&
  {Callanan}}]{2004ATel..366....1F}
{Filippenko}, A.~V., {Foley}, R.~J., {Callanan}, P.~J., 2004, The Astronomer's
  Telegram, 366, 1

\bibitem[{{Fox} \& {Kulkarni}(2004)}]{2004ATel..354....1F}
{Fox}, D.~B., {Kulkarni}, S.~R., 2004, The Astronomer's Telegram, 354, 1

\bibitem[{{Galloway} et~al.(2005){Galloway}, {Markwardt}, {Morgan},
  {Chakrabarty}, \& {Strohmayer}}]{2005ApJ...622L..45G}
{Galloway}, D.~K., {Markwardt}, C.~B., {Morgan}, E.~H., {Chakrabarty}, D.,
  {Strohmayer}, T.~E., 2005, \apjl, 622, L45

\bibitem[{{Heinke} et~al.(2003){Heinke}, {Grindlay}, {Lugger}, {Cohn},
  {Edmonds}, {Lloyd}, \& {Cool}}]{2003ApJ...598..501H}
{Heinke}, C.~O., {Grindlay}, J.~E., {Lugger}, P.~M., {Cohn}, H.~N., {Edmonds},
  P.~D., {Lloyd}, D.~A., {Cool}, A.~M., 2003, \apj, 598, 501

\bibitem[{{in't Zand} et~al.(2001){in't Zand}, {van Kerkwijk}, {Pooley},
  {Verbunt}, {Wijnands}, \& {Lewin}}]{2001ApJ...563L..41I}
{in't Zand}, J.~J.~M., {van Kerkwijk}, M.~H., {Pooley}, D., {Verbunt}, F.,
  {Wijnands}, R., {Lewin}, W.~H.~G., 2001, \apjl, 563, L41

\bibitem[{{Jonker} et~al.(2003){Jonker}, {M{\' e}ndez}, {Nelemans}, {Wijnands},
  \& {van der Klis}}]{2003MNRAS.341..823J}
{Jonker}, P.~G., {M{\' e}ndez}, M., {Nelemans}, G., {Wijnands}, R., {van der
  Klis}, M., 2003, \mnras, 341, 823

\bibitem[{{Jonker} et~al.(2004{\natexlab{a}}){Jonker}, {Galloway},
  {McClintock}, {Buxton}, {Garcia}, \& {Murray}}]{2004MNRAS.354..666J}
{Jonker}, P.~G., {Galloway}, D.~K., {McClintock}, J.~E., {Buxton}, M.,
  {Garcia}, M., {Murray}, S., 2004{\natexlab{a}}, \mnras, 354, 666

\bibitem[{{Jonker} et~al.(2004{\natexlab{b}}){Jonker}, {Wijnands}, \& {van der
  Klis}}]{2004MNRAS.349...94J}
{Jonker}, P.~G., {Wijnands}, R., {van der Klis}, M., 2004{\natexlab{b}},
  \mnras, 349, 94

\bibitem[{{Markwardt} et~al.(2004{\natexlab{a}}){Markwardt}, {Galloway},
  {Chakrabarty}, {Morgan}, \& {Strohmayer}}]{2004ATel..360....1M}
{Markwardt}, C.~B., {Galloway}, D.~K., {Chakrabarty}, D., {Morgan}, E.~H.,
  {Strohmayer}, T.~E., 2004{\natexlab{a}}, The Astronomer's Telegram, 360, 1

\bibitem[{{Markwardt} et~al.(2004{\natexlab{b}}){Markwardt}, {Swank}, \&
  {Strohmayer}}]{2004ATel..353....1M}
{Markwardt}, C.~B., {Swank}, J.~H., {Strohmayer}, T.~E., 2004{\natexlab{b}},
  The Astronomer's Telegram, 353, 1

\bibitem[{{Muno} et~al.(2003){Muno}, {Baganoff}, \&
  {Arabadjis}}]{2003ApJ...598..474M}
{Muno}, M.~P., {Baganoff}, F.~K., {Arabadjis}, J.~S., 2003, \apj, 598, 474

\bibitem[{{Nowak} et~al.(2004)}]{2004ATel..369....1N}
{Nowak}, M.~A., et~al., 2004, The Astronomer's Telegram, 369, 1

\bibitem[{{Pooley}(2004)}]{2004ATel..355....1P}
{Pooley}, G., 2004, The Astronomer's Telegram, 355, 1

\bibitem[{{Press} et~al.(1992){Press}, {Teukolsky}, {Vetterling}, \&
  {Flannery}}]{prteve1992}
{Press}, W.~H., {Teukolsky}, S.~A., {Vetterling}, W.~T., {Flannery}, B.~P.,
  1992, Numerical recipes in FORTRAN. The art of scientific computing,
  Cambridge: University Press, |c1992, 2nd ed.

\bibitem[{{Roelofs} et~al.(2004){Roelofs}, {Jonker}, {Steeghs}, {Torres}, \&
  {Nelemans}}]{2004ATel..356....1R}
{Roelofs}, G., {Jonker}, P.~G., {Steeghs}, D., {Torres}, M., {Nelemans}, G.,
  2004, The Astronomer's Telegram, 356, 1

\bibitem[{{Rutledge} et~al.(2001){Rutledge}, {Bildsten}, {Brown}, {Pavlov}, \&
  {Zavlin}}]{2001ApJ...551..921R}
{Rutledge}, R.~E., {Bildsten}, L., {Brown}, E.~F., {Pavlov}, G.~G., {Zavlin},
  V.~E., 2001, \apj, 551, 921

\bibitem[{{Rutledge} et~al.(2002{\natexlab{a}}){Rutledge}, {Bildsten}, {Brown},
  {Pavlov}, \& {Zavlin}}]{2002ApJ...577..346R}
{Rutledge}, R.~E., {Bildsten}, L., {Brown}, E.~F., {Pavlov}, G.~G., {Zavlin},
  V.~E., 2002{\natexlab{a}}, \apj, 577, 346

\bibitem[{{Rutledge} et~al.(2002{\natexlab{b}}){Rutledge}, {Bildsten}, {Brown},
  {Pavlov}, {Zavlin}, \& {Ushomirsky}}]{2002ApJ...580..413R}
{Rutledge}, R.~E., {Bildsten}, L., {Brown}, E.~F., {Pavlov}, G.~G., {Zavlin},
  V.~E., {Ushomirsky}, G., 2002{\natexlab{b}}, \apj, 580, 413

\bibitem[{{Shaw} et~al.(2005)}]{2005A&A...432L..13S}
{Shaw}, S.~E., et~al., 2005, \aap, 432, L13

\bibitem[{{Steeghs} et~al.(2004){Steeghs}, {Blake}, {Bloom}, {Torres},
  {Jonker}, \& {Starr}}]{2004ATel..363....1S}
{Steeghs}, D., {Blake}, C., {Bloom}, J.~S., {Torres}, M.~A.~P., {Jonker},
  P.~G., {Starr}, D., 2004, The Astronomer's Telegram, 363, 1

\bibitem[{{Wijnands} \& {van der Klis}(1998)}]{1998Natur.394..344W}
{Wijnands}, R., {van der Klis}, M., 1998, \nat, 394, 344

\bibitem[{{Wijnands} \& {van der Klis}(1999)}]{1999ApJ...514..939W}
{Wijnands}, R., {van der Klis}, M., 1999, \apj, 514, 939

\bibitem[{{Wijnands} et~al.(2002){Wijnands}, {Guainazzi}, {van der Klis}, \&
  {M{\' e}ndez}}]{2002ApJ...573L..45W}
{Wijnands}, R., {Guainazzi}, M., {van der Klis}, M., {M{\' e}ndez}, M., 2002,
  \apjl, 573, L45

\bibitem[{{Wijnands} et~al.(2004){Wijnands}, {Homan}, {Miller}, \&
  {Lewin}}]{2004wijninpress}
{Wijnands}, R., {Homan}, J., {Miller}, J.~M., {Lewin}, W.~H.~G., 2004, \apjl,
  606, L61

\bibitem[{{Wijnands} et~al.(2005){Wijnands}, {Homan}, {Heinke}, {Miller}, \&
  {Lewin}}]{2005ApJ...619..492W}
{Wijnands}, R., {Homan}, J., {Heinke}, C.~O., {Miller}, J.~M., {Lewin},
  W.~H.~G., 2005, \apj, 619, 492

\end{thebibliography}
\end{document}